\documentclass[conference]{IEEEtran}
\IEEEoverridecommandlockouts
\usepackage{cite}
\usepackage{amsmath,amssymb,amsfonts,enumerate}
\usepackage{algorithmic}
\usepackage{graphicx} \graphicspath{{outputs/}}
\usepackage{textcomp}
\usepackage[bookmarks=false, draft]{hyperref}
\def\BibTeX{{\rm B\kern-.05em{\sc i\kern-.025em b}\kern-.08em
    T\kern-.1667em\lower.7ex\hbox{E}\kern-.125emX}}
\begin{document}

\IEEEpubid{978-1-7281-2822-1/20/\$31.00 ̃\copyright ̃2020 IEEE \hspace*{40.5em} PMAPS 2020}

\title{\textit{Importance subsampling} for power system planning under multi-year demand and weather uncertainty\\
\thanks{This work was supported by the United Kingdom Engineering and Physical Sciences Research Council (EPSRC) Mathematics of Planet Earth Centre for Doctoral Training, grant number EP/L016613/1.}
}

\author{\IEEEauthorblockN{Adriaan P. Hilbers}
\IEEEauthorblockA{\textit{Department of Mathematics} \\
\textit{Imperial College London}\\
London, United Kingdom\\
ORCID: 0000-0002-9882-9479
 }
\and
\IEEEauthorblockN{David J. Brayshaw}
\IEEEauthorblockA{\textit{Department of Meteorology} \\
\textit{University of Reading}\\
Reading, United Kingdom\\
ORCID: 0000-0002-3927-4362
 }
\and
\IEEEauthorblockN{Axel Gandy}
\IEEEauthorblockA{\textit{Department of Mathematics} \\
\textit{Imperial College London}\\
London, United Kingdom\\
ORCID: 0000-0002-6777-0451
 }
}

\maketitle

\begin{abstract}
This paper introduces a generalised version of \textit{importance subsampling} for time series reduction/aggregation in optimisation-based power system planning models. Recent studies indicate that reliably determining optimal electricity (investment) strategy under climate variability requires the consideration of multiple years of demand and weather data. However, solving planning models over long simulation lengths is typically computationally unfeasible, and established time series reduction approaches induce significant errors.  The \textit{importance subsampling} method reliably estimates long-term planning model outputs at greatly reduced computational cost, allowing the consideration of multi-decadal samples. The key innovation is a systematic identification and preservation of relevant extreme events in modeling subsamples. Simulation studies on generation and transmission expansion planning models illustrate the method's enhanced performance over established ``representative days'' clustering approaches. The models, data and sample code are made available as open-source software.
\end{abstract}

\begin{IEEEkeywords}
Power system modeling, weather, representative day, subsampling, temporal aggregation
\end{IEEEkeywords}

\section{Introduction}
\label{sec:introduction}

\subsection{Demand and weather uncertainty and computational complexity in power system planning models}
\label{sec:introduction:demand_weather_uncertainty}

Many \textit{power system planning models} estimate optimal electricity investment strategy by minimising a cost function given techno-economic assumptions and constraints \cite{bazmi_2011}. The use of demand and weather time series inputs (e.g. hourly load levels and wind speeds) induces forward propagated \textit{demand and weather uncertainty} in model outputs, particularly when variable renewable (VR) generation such as solar and wind is considered \cite{hilbers_2020}. This is because an input time series may be viewed as a sample from some underlying demand and weather distribution, and a different sample, leading to different outputs, is equally valid.

Recent studies indicate that inter-annual climate/weather variability induces considerable output uncertainty in planning models. For example, optimal system design may depend strongly on what year of data is used \cite{bloomfield_2016, zeyringer_2018, staffel_2018, collins_2018, bothwell_2018}, with some optimal capacities differing as much as 80\% \cite{pfenninger_2017, hilbers_2019}. Hence, to reliably determine long-term optimal electricity strategy, multiple years of demand and weather data should be considered.

At the same time, using long samples of time series input data is unfeasible in many power system planning models due to the computational complexity in solving the associated optimisation problem \cite{ringkjob_2018}. This effect is amplified since accurately modeling VR generation requires a high temporal and spatial resolution, precluding an increase in step length to reduce the number of time steps \cite{poncelet_2016, collins_2017, kotzur_2018}.

\subsection{Time series reduction and subsampling}
\label{sec:introduction:time_series_reduction}

\IEEEpubidadjcol

Power system modelers employ various schemes to reduce temporal complexity, as discussed in detail in \cite{hoffmann_2020}. A popular approach involves subsampling, often by clustering, into a smaller number of \textit{representative periods} (typically days) that seek to encode longer time series in reduced form \cite{fitiwi_2015, nahmmaccher_2016, hartel_2017, poncelet_2017, tejada_2018}.

Most such reduction approaches are what the authors of \cite{hoffmann_2020} call \textit{a priori} (or \textit{input-based}) in that they depend only on the input time series but not on the underlying power system model. For example, clustering the same demand and weather data generates the same representative days (ignoring any randomness in the clustering algorithm) irrespective of whether these days are subsequently used in a highly renewable system or a system with only fossil fuel generation. In constrast, \textit{a posteriori} methods use knowledge of the model's structure or an estimate of its results. As an example, in \cite{sun_2019}, the authors run a planning model on each day individually and aggregate days by clustering (with additional modifications) the vectors of associated investment costs (model outputs). In this way, they use a subsample tailored to the model in question.

Many time series aggregation methods, such as clustering, replace individual periods/days by a group average. This ``smoothes out'' extremes and may lead to (1) overly optimistic predictions of system costs and (2) system design estimates that are not robust to such events. For example, such approaches may underestimate the amount of required backup generation capacity in events with high demand but low renewable generation, leading to systems with a high risk of blackouts \cite{hoffmann_2020}. For this reason, users sometimes make \textit{a priori} heuristic adjustments such as including the maximum demand day \cite{pfenninger_2017}.

\subsection{This paper's contribution}
\label{sec:introduction:contribution}

This paper generalises significantly a version of the \textit{importance subsampling} method for time series reduction/aggregation originally introduced in \cite{hilbers_2019}. The contribution is as follows. Section \ref{sec:introduction:demand_weather_uncertainty} indicates that long demand and weather samples should be considered for robust power system planning under climate/weather variability, but that this is computationally challenging. Furthermore, time series reduction methods that remove extremes (Section \ref{sec:introduction:time_series_reduction}) lead to inaccurate estimates of optimal electricity strategy. The \textit{importance subsampling} method systematically identifies such extremes in the full time series and includes them in a subsample, allowing reliable estimation of optimal multi-year electricity strategy at greatly reduced computational cost. A simulation study demonstrates the method's improved performance over established clustering methods. The models, data and sample code for the method are available at \cite{github_iss}.

This paper extends the approach introduced in \cite{hilbers_2019} to almost general cost-minimisation based power system planning models. The requirements for the original method (analytical formula for generation cost, no link between time steps) restricted applications to models solvable using the (net) load duration curve. These stipulations are all relaxed. Furthermore, the generalised method allows, in addition to sampling individual time steps, the sampling of representative periods such as days or weeks (hence allowing for consideration of intra-period temporal dependencies) and uses clustering to leverage off their extensive use in power system applications.

This paper is structured as follows. Section \ref{sec:importance_subsampling} introduces the method and the intuition behind its working. Section \ref{sec:simulations} provides a case study of the method's appliction to two generation and expansion planning (GTEP) models. Section \ref{sec:discussion_conclucions} discusses results and recommends extensions.

\section{Importance subsampling}
\label{sec:importance_subsampling}

\subsection{Overview and intuition}
\label{sec:importance_subsampling:overview}

This paper considers planning models that determine optimal system design $\mathbf{D}_{\text{opt}}$ (e.g. installed capacities of generation and transmission technologies) given some (possibly weighted) demand and weather data. The model $\text{PSM}_\text{plan}$ is viewed as a function from a sample $\mathcal{T}$ (of length $|\mathcal{T}|$) of demand and weather data to $\mathbf{D}_{\text{opt}}$:
\begin{equation}
  \label{eq:psm:planning_overview}
  \mathbf{D}_\text{opt} = \text{PSM}_{\text{plan}}(\mathcal{T}).
\end{equation}
The model can also run in \textit{operational} mode, in which a design $\mathbf{D}$ is specified and the optimal system operation, typically with the lowest generation cost, is determined:
\begin{equation}
  \label{eq:psm:operate_overview}
  (O_1, \ldots, O_{|\mathcal{T}|}) = \text{PSM}_\text{operate}(\mathcal{T}|\mathbf{D}).
\end{equation}
Here, $O_{t}$ is the power system operation (generation and transmission levels) at time $t$, for each time step $t=1, \ldots, |\mathcal{T}|$. In many planning models (e.g. ones that seek the design minimising the sum of build costs and subsequent minimum generation costs), the operational model is solved as a subproblem. However, this is not necessary for the proposed method.

The setup is then as follows. Suppose $\mathcal{T}$ is a contiguous sample of demand and weather data and the associated optimal system design $\mathbf{D}_{\text{opt}}$, as defined by (\ref{eq:psm:planning_overview}), is desired. If $\mathcal{T}$ is long, solving the optimisation problem to determine $\mathbf{D}_\text{opt}$ may be computationally unfeasible. The \textit{importance subsampling} method instead estimates $\mathbf{D}_\text{opt}$ by running the planning model across a shorter (weighted) subsample of time series data $\mathcal{S}$. An estimate of $\mathbf{D}_\text{opt}$ in (\ref{eq:psm:planning_overview}) is hence provided by
\begin{equation}
  \label{eq:psm:iss_est}
  \mathbf{D}_\text{opt}^{\mathcal{S}} := \text{PSM}_{\text{plan}}(\mathcal{S}).
\end{equation}
The goal is that $\mathbf{D}_\text{opt}^{\mathcal{S}} \approx \mathbf{D}_\text{opt}$ with $|\mathcal{S}| < |\mathcal{T}|$, so that optimal system design is reliably estimated at reduced computational cost. The key is hence how the sample $\mathcal{S}$ is created.

The \textit{importance subsampling} method works by explicitly including a selection of ``extreme'' events in a modeling subsample. In this paper, ``extreme'' refers to time steps in which meeting demand requires a large amount of generation or transmission capacity. In many planning models, omitting such extremes leads to inaccurate estimates of optimal system design (e.g. underestimation of required backup generation capacity, see Section \ref{sec:introduction:demand_weather_uncertainty}). It is hence essential that such events are identified and preserved when subsampling. At the same time, identifying them \textit{a priori} is difficult since which time steps are ``extreme'' may depend on the system design being estimated. As a simple example, in a model containing an unknown wind generation capacity, it is unclear whether a 55MW demand, 0.1 wind capacity factor time step is more ``extreme'' (in the sense of requiring more backup capacity) than one with 60MW demand and 0.2 wind capacity factor.

\textit{Importance subsampling} systematically identifies relevant extremes using their \textit{importance}, calculated using the user-defined function $\text{IMP}$ (see Section \ref{sec:importance_subsampling:steps}). This function should be a proxy for how ``extreme'' a time step is. In many applications (e.g. Section \ref{sec:simulations}), a time step's \textit{generation cost} performs well since expensive measures (e.g. peaking plants or load curtailment) are activated only in scenarios when there is otherwise a supply shortage. Other choices are possible; electricity price is another candidate in models with such a notion. In practice, expert knowledge of the system should motivate a choice of \textit{importance} function.

Determining each time step's \textit{importance} generally requires the associated generation and transmission levels, determined by running the model in \textit{operational} mode with some fixed design. Ideally, \textit{importance} levels are calculated using the true optimal system design, but this is the output being estimated and hence unknown. For this reason, a preliminary planning model run with a standard subsampling scheme is used for a rough estimate of optimal system design. This is subsequently used to determine the \textit{importance} of each time step in the full time series.

At a high level, the approach works as follows (with correspondence to Section \ref{sec:importance_subsampling:steps} shown in parantheses):
\begin{enumerate}[i)]
\item Estimate optimal system design through model run with standard subsampling (\textit{a priori}) procedure (steps 1-2).
\item Using this estimate, calculate \textit{importance} for each time step in full time series. In general, this requires running the model in operational mode (steps 3-4).
\item Create \textit{importance subsample} by subsampling again while explicitly preserving the relevant extreme events (those with high \textit{importance}) (step 5).
\end{enumerate}
A model is subsequently run with the \textit{importance subsample} $\mathcal{S}$ to estimate optimal system design.

\subsection{Algorithm}
\label{sec:importance_subsampling:steps}

\noindent \hrulefill \\
\textbf{Algorithm: \textit{Importance subsampling}} \\
\textbf{Inputs}:
\begin{itemize}
\item $\mathcal{T}$: full time series to sample from, of length $|\mathcal{T}|$
\item $n_d$: number of days to sample
\item $n_{d_e}$: number of ``extreme'' days to sample
\end{itemize}
\textbf{Steps}:
\begin{enumerate}
\item Cluster full time series into $n_d$ weighted representative days to create $\mathcal{C}$.
\item Run planning model to determine $\mathbf{D}_\text{opt}^\mathcal{C}$, the cluster estimates of optimal system design:
  \begin{equation}
    \label{eq:iss_run_s2}
    \mathbf{D}_\text{opt}^\mathcal{C} = \text{PSM}_{\text{plan}}(\mathcal{C}).
  \end{equation}
\item Determine system operation $(O_1^{\mathcal{C}}, \ldots, O_{|\mathcal{T}|}^{\mathcal{C}})$ (generation and transmission levels at each time in $\mathcal{T}$) by running model in operational mode with design $\mathbf{D}_\text{opt}^\mathcal{C}$:
  \begin{equation}
    \label{eq:iss_run_forward_s1}
    (O_1^{\mathcal{C}}, \ldots, O_{|\mathcal{T}|}^{\mathcal{C}}) = \text{PSM}_\text{operate}(\mathcal{T}|\mathbf{D}_\text{opt}^\mathcal{C}).
  \end{equation}
\item Assign each time step its \textit{importance}:
  \begin{equation}
    \label{eq:iss_importance}
    \text{imp}_t = \text{IMP}(O_t^{\mathcal{C}}) \quad \forall t \in \mathcal{T}.
  \end{equation}
for user-defined $\text{IMP}$ function (see Section \ref{sec:importance_subsampling:overview}).
\item Create $n_d$-day \textit{importance subsample} $\mathcal{S}$ containing:
  \begin{itemize}
  \item $n_{d_e}$ most ``extreme'' days in $\mathcal{T}$ as ranked by the \textit{importance} of their highest time step.
  \item $n_d - n_{d_e}$ days obtained by clustering the remainder of the full time series. 
  \end{itemize}
Weight days to account for their proportion across the full time series $\mathcal{T}$ (see Section \ref{sec:importance_subsampling:intuition_remarks}).
\end{enumerate}
\textbf{Outputs}:
\begin{itemize}
\item $\mathcal{S}$: weighted \textit{importance subsample}
\end{itemize}
\noindent \hrulefill

\subsection{Remarks}
\label{sec:importance_subsampling:intuition_remarks}

The proposed method is an \textit{a posteriori} time series aggregation scheme that customises the subsample to the model. Most established time series reduction methods (Section \ref{sec:introduction:demand_weather_uncertainty}) are \textit{a priori} schemes which do not use knowledge of the underlying model. This corresponds to following just step i in Section \ref{sec:importance_subsampling:overview} (steps 1-2 in Section \ref{sec:importance_subsampling:steps}) and using $\mathbf{D}_{\text{opt}}^{\mathcal{C}}$ as the final estimate. \textit{Importance subsampling} instead uses this estimate to create an additional customised sample to provide a second, more accurate, estimate of optimal system design.

Estimating optimal system design using \textit{importance subsampling} requires three model runs (with correspondence to Section \ref{sec:importance_subsampling:steps} shown in parantheses):
\begin{enumerate}
\item Preliminary planning model run on $n_d$ days (step 2)
\item Operational model run with fixed capacities on full time series (step 3)
\item Planning model run on $n_d$-day \textit{importance subsample}.
\end{enumerate}
The first and third runs are planning models that determine optimal system design. These models usually also determine the associated optimal generation levels as part of the optimisation problem, but this output is ignored. In contrast, the second run determines just the system's operation with a fixed design. Operational models' solution times typically scale linearly in the number of time steps and are much shorter than for planning models \cite{pfenninger_2018}. For this reason, a planning model run with \textit{importance subsample} requires roughly twice the computational time as a single planning model run across $n_d$ days.

The \textit{importance subsample} is weighted to account for the relative proportions of days throughout the full time series. For example, the 30 most extreme days in a 30-year sample represent one day a year, but will represent 30 days a year if all included a 1-year subsample. The time steps are hence weighted to avoid the estimated optimal system design being over-engineered for extreme events. The $n_d - n_{d_e}$ non-extreme days are weighted according to their cluster sizes.

The clustering/subsampling algorithms used in steps 1 and 5 of Section \ref{sec:importance_subsampling:steps} are customisable. Furthermore, periods other than days may be sampled. The advantage of sampling longer periods (e.g. weeks) is that it respects long-term temporal dependencies such as storage to a greater degree. The advantage of sampling shorter blocks is that the total number of periods in a fixed-length subsample is larger, meaning a higher diversity of possible demand and weather scenarios is considered. Hence, the choice of subsample blocks is a tradeoff between a high diversity of scenarios and a low distortion of time step chronology. In models without temporal dependencies, randomly sampling individual time steps (without any clustering) typically gives the best results, but longer periods should be selected in models with ramping or storage constraints.

\section{Simulation studies}
\label{sec:simulations}

\subsection{Overview}
\label{sec:simulations:overview}

This section examines the efficacy of the proposed method on sample power system planning models. This is done by comparing optimal system design across \textit{importance subsamples} with that across the full time series from which samples are drawn. As a benchmark, the model is also run across samples generated by $k$-medoids clustering, representing the current industry standard. The reason for choosing $k$-medoids clustering (where the real day closest to the cluster mean is used as the representative day, instead of the mean itself) is its widespread use in previous studies and that it typically performs better than $k$-means \cite{hoffmann_2020}, which smooths out extremes by using the cluster mean.

Two metrics indicating the ``degree of suboptimality'' of estimated system design (compared with the true optimum across the full time series) are also calculated:
\begin{itemize}
\item \textit{Peak capacity shortage}: maximum value of demand minus available generation capacity, in MW.
\item \textit{Energy unserved}: (annualised) demand impossible to meet with generation capacity, in MWh/yr. Can be turned into monetary value via a value of lost load.
\end{itemize}
These metrics are determined by running a model, with design estimated across subsamples, in \textit{operational} mode across the full time series.

\subsection{Power system planning models}
\label{sec:simulations:models}

\begin{figure}
  \small
  \setlength{\tabcolsep}{0.3em}
  \begin{tabular}{c}
    \includegraphics[scale=0.43, trim=10 10 10 10, clip]{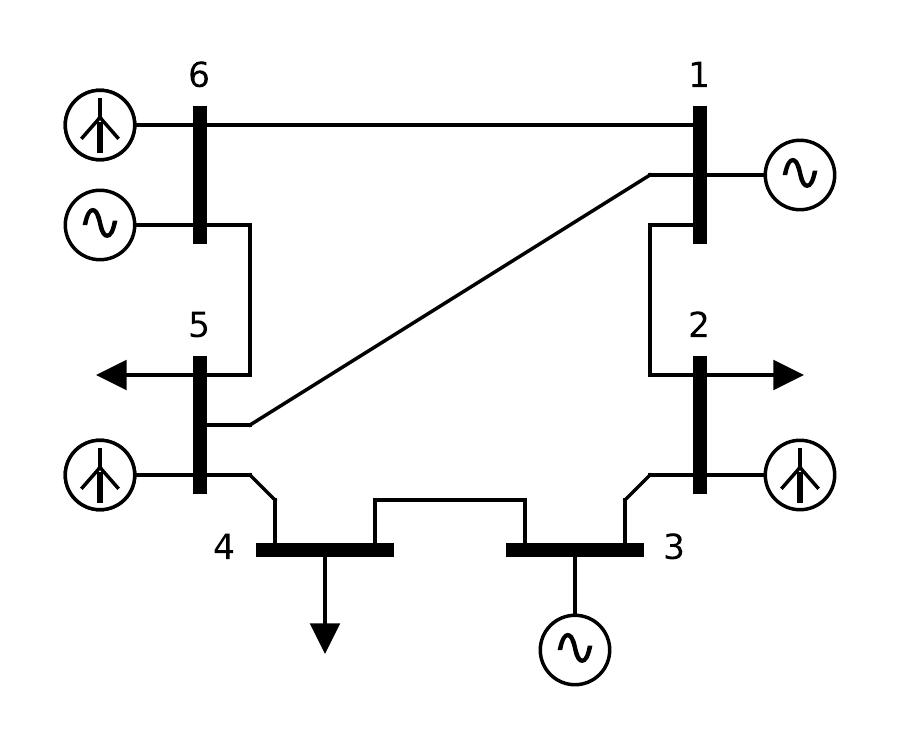}
  \end{tabular}
  \begin{tabular}{r  c}
  Bus & Demand/candidate generation \\ \hline
  1 & baseload, peaking \\
  2 & demand (DE), wind (DE) \\
  3 & baseload, peaking \\
  4 & demand (FR) \\
  5 & demand (UK), wind (UK) \\
  6 & baseload, peaking, wind (ES) \\ \hline
\end{tabular}
  \caption{Power system planning model configuration. Time-varying demand is met at buses 2, 4 and 5. Thermal generation (baseload or peaking) may be built at buses 1, 3 and 6, while wind power may be installed at buses 2, 5 and 6. Time series input data are aggregated demand and wind levels for Germany (DE), France (FR), United Kingdom (UK) and Spain (ES).}
  \label{fig:simulation:model_diagram}
\end{figure}

The method is tested on two (generation and transmission expansion) planning models with topologies equal to a renewables-containing version of the \textit{IEEE 6-bus system} originally introduced in \cite{kamalinia_2010, kamalinia_2011}. Optimal system design (generation and transmission capacities across the grid) is determined by minimising the total system cost (sum of install and generation costs) given hourly demand levels and wind capacity factors. The model determines the optimal capacity and generation mix across \textit{baseload}, \textit{peaking} and \textit{wind} technologies. Time series input data consists of hourly demand levels and wind capacity factors in European countries, introduced in \cite{bloomfield_2019, bloomfield_MERRA2}. An overview of the model's topology is provided in Fig. \ref{fig:simulation:model_diagram}. The models and time series data are publicly available at \cite{github_renewable_test_psms}, where their mathematical formulations are described.

The two models differ in the complexity of their constraints. The first, the \textit{LP} model, is a linear program (LP) in which technologies may be installed to any nonnegative capacity and there is no intertemporal dependence between time steps. In the \textit{MILP} model, baseload capacity is constrained to blocks of 3GW and a ramp rate of 20\%/hr, making the model a \textit{mixed-integer linear program} (MILP) with ramping constraints.

\subsection{Simulation setup}
\label{sec:simulations:setup}

\begin{table}
  \caption{Simulations in Section \ref{sec:simulations:results}. $n_{d_e}$ is the number of ``extreme'' days included in subsamples. For \textit{importance subsampling}, solution times are disaggregated into first planning run, operational run and second planning run.}
\centering
\footnotesize
\setlength{\tabcolsep}{0.5em}
\begin{tabular}{r | c | c | c | c}
  \multicolumn{3}{c|}{Time series data} & \multicolumn{2}{c}{Mean solution requirements} \\
  & length $n_d$ & $n_{d_e}$ & Time & Memory \\
  Subsampling & (days) & (days) & (min) & (MB) \\ \hline \hline
  \multicolumn{5}{c}{} \\
  \multicolumn{5}{c}{\textbf{LP model}, full time series 2008-2017} \\ \hline \hline
  None (target) & 3650 & - & 340 & 15121 \\ \hline
  $k$-medoids & 48 & - & 0.5 & 393 \\
  \textit{importance} & 48 & 16 & 0.5+11+0.5 & 742 \\ \hline
  $k$-medoids & 90 & - & 1 & 567 \\
  \textit{importance} & 90 & 30 & 1+11+1 & 753 \\ \hline \hline
  \multicolumn{5}{c}{} \\
  \multicolumn{5}{c}{\textbf{MILP model}, full time series 2017} \\ \hline \hline
  None (target) & 365 & - & 2926 & 1598 \\ \hline
  $k$-medoids & 24 & - & 2 & 270 \\
  \textit{importance} & 24 & 8 & 2+2+2 & 823 \\ \hline
  $k$-medoids & 48 & - & 9 & 362 \\
  \textit{importance} & 48 & 16 & 9+2+9 & 841 \\ \hline \hline
\end{tabular} \vspace{1em}
  \label{tab:simulations}
\end{table}

Details for the simulations are provided in Table \ref{tab:simulations}. For the LP model, optimal design across 2008-2017 serves as the target for simulations with $n_d$=48 and 90-day subsamples. For the MILP model, the integer constraints impose (much) longer solution times and the optimal design across 2017 is estimated using subsamples of length $n_d$=24 and 48 days. Since both $k$-medoids clustering and \textit{importance subsampling} contains randomness from the initialisation of cluster means, the distribution of model outputs across 100 subsamples is determined. Computational times are based on model runs created in the open-source energy modeling framework \textsl{Calliope} \cite{pfenninger_2018} and solved using the \textsl{cbc} \cite{cbc} solver on a 2.7GHz Intel Core i5-5257U processor with 8GB of RAM (with some additional ``swap'' memory available).

The implementation of the \textit{importance subsampling} method (Section \ref{sec:importance_subsampling:steps}) is as follows. $k$-medoids clustering is used in steps 1 and 5. Each time series is normalised to lie between 0 and 1 (so that demand levels and wind capacity factors can be compared fairly), but no additional adjustments are made. For the \textit{importance} function in step 4, a time step's \textit{generation cost} (sum of each technology's generation cost times generation level) is used since expensive technologies with high generation cost (e.g. peaking plants) are used only in ``extreme'' scenarios when blackouts would occur otherwise. For $n_{d_e}$, the number of ``extreme'' days containing the highest generation costs, a value of $n_{d_e} = \frac{1}{3}n_d$ is used as this gives good results in practice, balancing the need to capture a range of ``extreme'' events while leaving enough of the sample to accurately approximate the more typical days.

\subsection{Results}
\label{sec:simulations:results}

\begin{figure}
  \centering
  \includegraphics[scale=0.5, trim=10 146 10 0, clip]{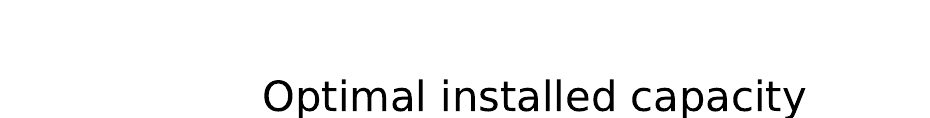}
  \includegraphics[scale=0.5, trim=10 146 10 0, clip]{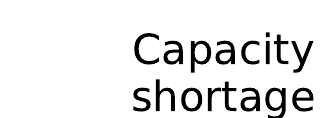}
  \includegraphics[scale=0.5, trim=10 146 10 0, clip]{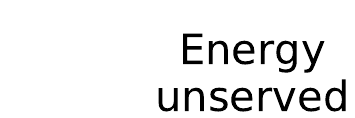} \vspace{0.5em} \\
  \textbf{LP model}, full time series 2008-2017 \\ \vspace{0.5em}
  (a) $n_d$=48-day subsample, $n_{d_e}$=16 ``extreme'' days \\ \vspace{0.5em}
  \includegraphics[scale=0.5, trim=10 10 10 34, clip]{LP/figs/caps_days_48.pdf}
  \includegraphics[scale=0.5, trim=10 10 10 34, clip]{LP_operate/figs/peak_unmet_days_48.pdf}
  \includegraphics[scale=0.5, trim=10 10 10 34, clip]{LP_operate/figs/gen_unmet_days_48.pdf} \vspace{1.0em} \\
  (b) $n_d$=90-day subsample, $n_{d_e}$=30 ``extreme'' days \\ \vspace{0.5em}
  \includegraphics[scale=0.5, trim=10 10 10 34, clip]{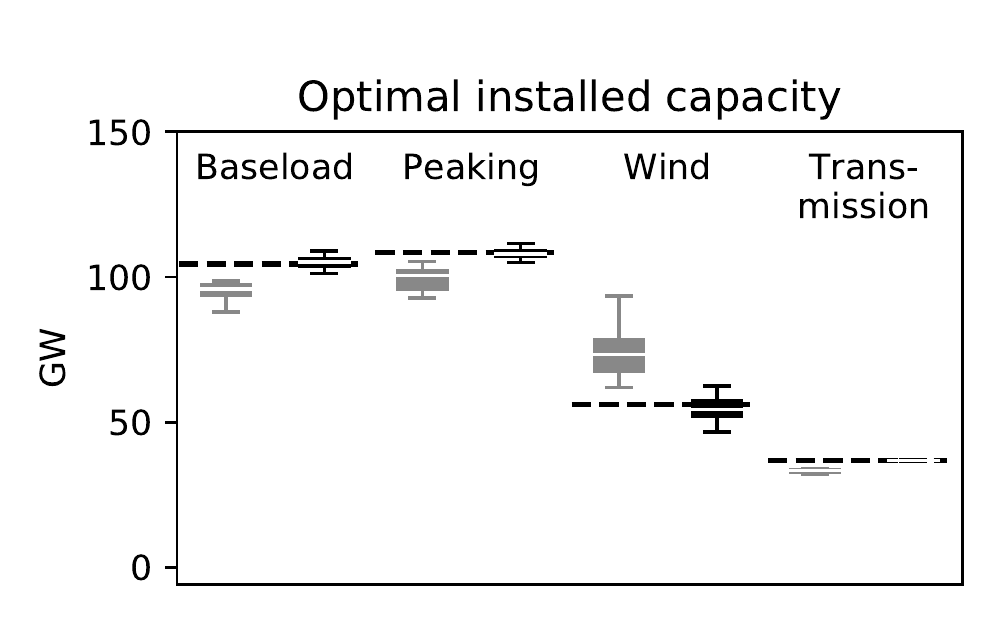}
  \includegraphics[scale=0.5, trim=10 10 10 34, clip]{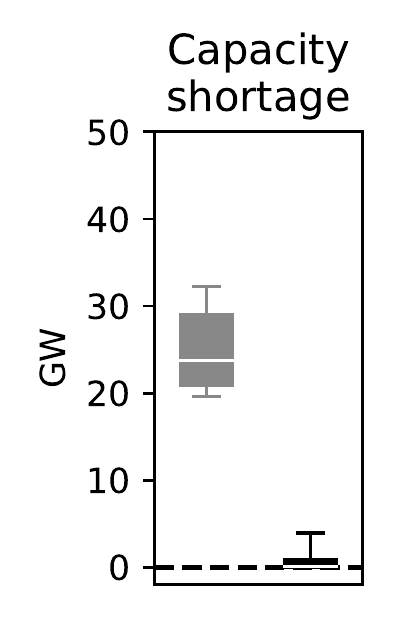}
  \includegraphics[scale=0.5, trim=10 10 10 34, clip]{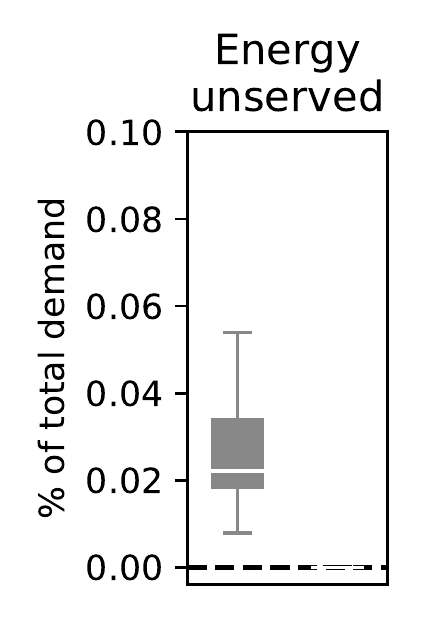} \vspace{1.0em} \\
  \textbf{MILP model}, full time series 2017 \\ \vspace{0.5em}
  (c) $n_d$=24-day subsample, $n_{d_e}$=8 ``extreme'' days \\ \vspace{0.5em}
  \includegraphics[scale=0.5, trim=10 10 10 34, clip]{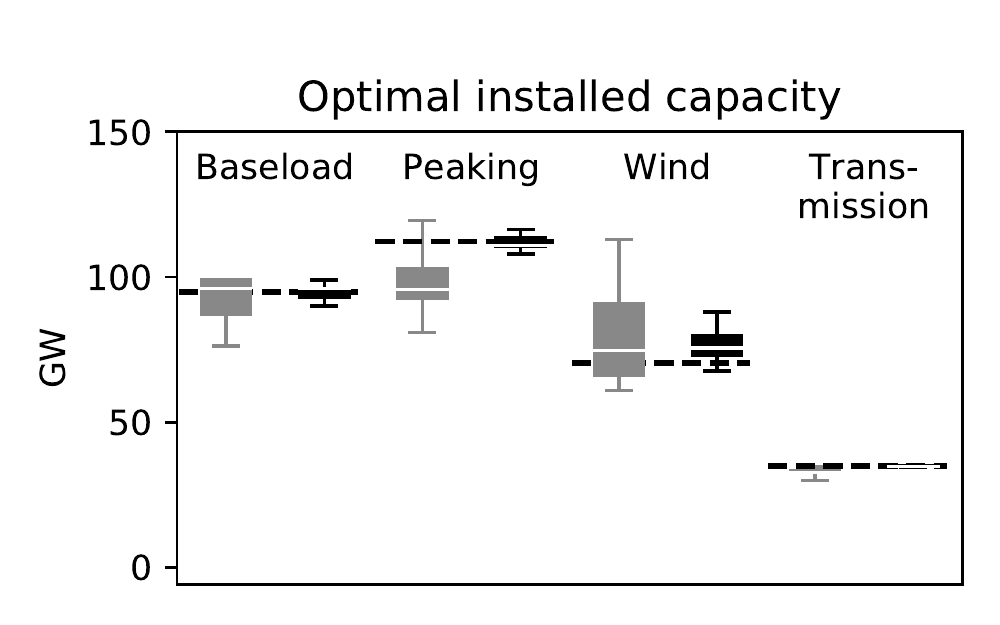}
  \includegraphics[scale=0.5, trim=10 10 10 34, clip]{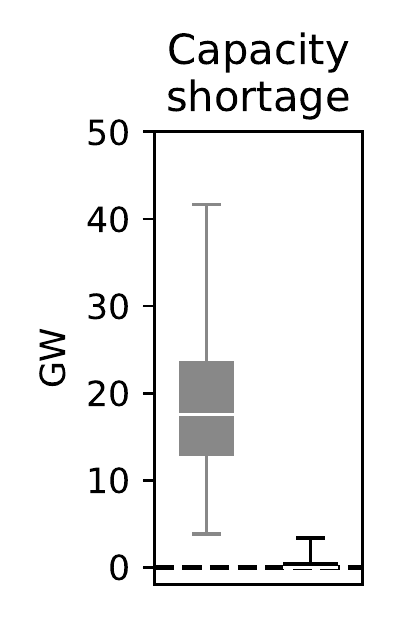}
  \includegraphics[scale=0.5, trim=10 10 10 34, clip]{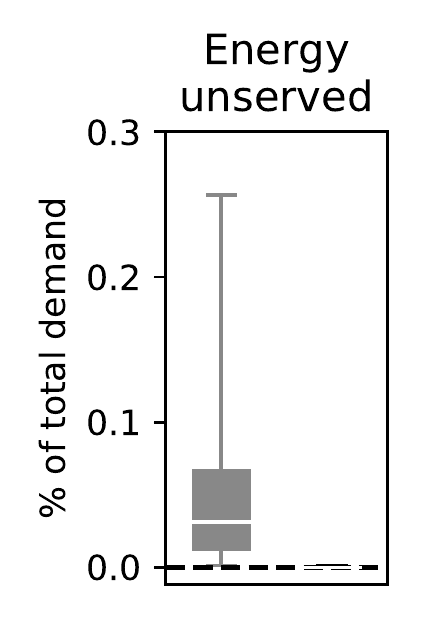} \vspace{1.0em} \\
  (d) $n_d$=48-day subsample, $n_{d_e}$=16 ``extreme'' days \\ \vspace{0.5em}
  \includegraphics[scale=0.5, trim=10 10 10 34, clip]{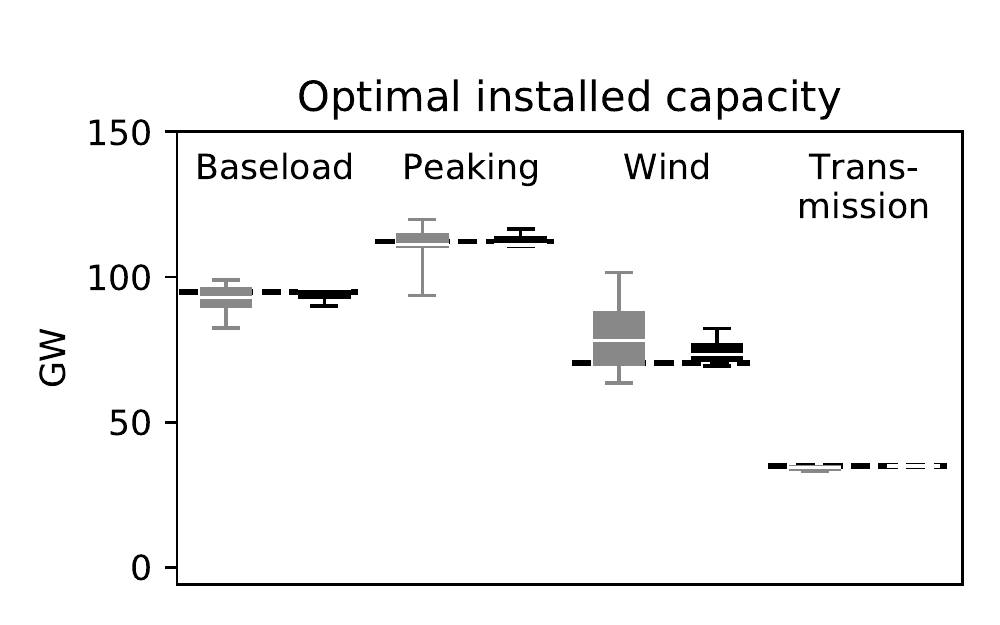}
  \includegraphics[scale=0.5, trim=10 10 10 34, clip]{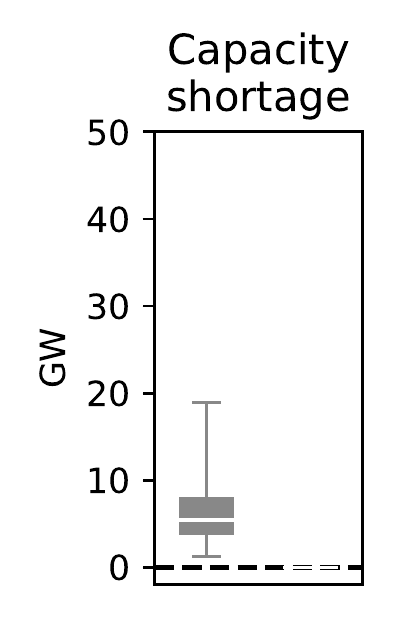}
  \includegraphics[scale=0.5, trim=10 10 10 34, clip]{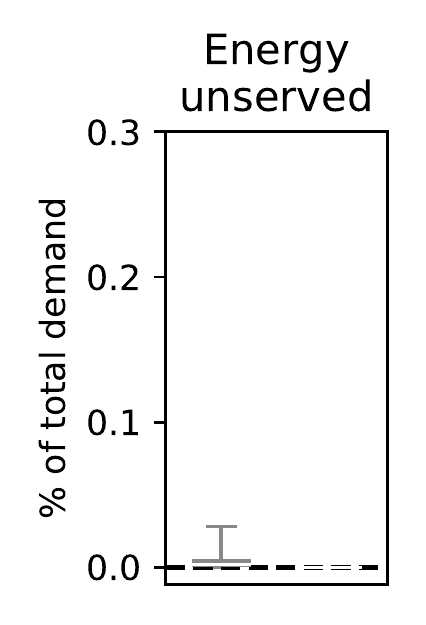} \vspace{1.0em} \\
  \includegraphics[scale=0.55, trim=0 0 0 0, clip]{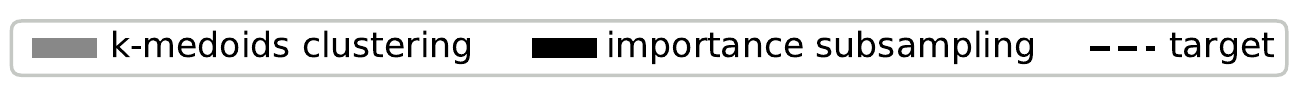}  \\
  \caption{Distribution of estimated optimal capacities and suboptimality metrics (peak capacity shortage, energy unserved) across 100 samples generated by $k$-medoids clustering and \textit{importance subsampling}. The box-and-whiskers show the 2.5\%, 25\%, median, 75\% and 97.5\% percentiles. The optima across the full time series (targets) are shown as a dashed black line.}
  \label{fig:results}
\end{figure}

Fig. \ref{fig:results} shows the distribution of estimated optimal capacities and suboptimality metrics (peak capacity shortage, energy unserved) across the test simulations. For the same plots among the full range of model outputs (at regional level) and for a larger number of sample sizes, including sampling individual hourly time steps instead of days, visit \cite{github_iss}. \textit{Importance subsampling} into representative days generally has superior performance to $k$-medoids clustering, with more accurate estimates of optimal system design, lower levels of capacity shortage and energy unserved, and lower variability in model outputs. For example, $k$-medoids clustering usually underestimates (overestimates) optimal peaking (wind) capacity, but such biases are not present for \textit{importance subsampling}. As a result, levels of capacity shortage and unmet demand across the full time series (rightmost two columns) are lower (almost 0) for \textit{importance subsampling} than for $k$-medoids. A system designed based on 48 representative days $k$-medoids clustered from 10 years (Fig. \ref{fig:simulation:model_diagram}(a)) fails to meet 0.05\% of demand on average, compared to virtually none for a system designed using \textit{importance subsampling}. This has financial implications; assigning a value of lost load of $\pounds$6000/MWh \cite{elexon_2015} implies an increase in system cost of around 1\% per each 0.01\% of demand that is unmet. In summary, \textit{importance subsampling} provides estimates of optimal system design that are both more accurate (estimated optimal capacities are closer to the true optimum across the full time series) and more robust to extreme events (much lower levels of capacity shortage and unserved energy).

\subsection{Discussion}
\label{sec:simulations:discussion}

The superior performance of \textit{importance subsampling} is a direct result of the method's explicit preservation of extreme days. By design, $k$-medoids clustering is equivalent to using $\mathcal{C}$ in step 1 (Section \ref{sec:importance_subsampling:steps}) to generate the final estimates of optimal system design. Hence, the improved performance of \textit{importance subsampling} comes only from steps 2-5, in which extreme days are identified and included. It is to be expected that the failure to include these events results in the underestimation (overestimation) of peaking (wind) capacity observed in Fig. \ref{fig:simulation:model_diagram}; samples not including high demand, low wind events underestimate the amount of required backup generation capacity while overestimating the reliability of wind power.

The performance gap of \textit{importance subsampling} over $k$-medoids clustering appears more strongly when subsampling from longer time series, as seen by comparing Fig. \ref{fig:results}(a) and (d). This is because representing a shorter time series (with a smaller range of extreme events) using e.g. clustering is easier than doing the same (with equal subsample size) for a longer, more diverse time series. Hence, the added value of \textit{importance subsampling} is expected to grow as longer time series (e.g. multiple decades) are sampled from.

The \textit{importance subsampling} method can estimate model outputs from simulations much longer than those used here. For example, it is possible to estimate a 25-year output by a 1-year subsample. The use of shorter samples is only to make validation of the method possible by comparing with the ``ground truth'' determined by a brute force model run on the full time series. The use of short samples in this manner has also distorted the computational expense. In most realistic planning models, computational expense for \textit{importance subsampling} is dominated by the two planning runs. This is because solution times for the operational model scale linearly in the number of time steps while that for a planning run scales faster. For example, for the MILP model, a 1-year estimate of 25-year optimal design requires around 48 hours for each planning run but just 48 minutes for the operational run across the full 25 years.

\section{Conclusions}
\label{sec:discussion_conclucions}

This paper's main contribution is a significant generalisation of the \textit{importance subsampling} method. This approach can reliably estimate long-term planning model outputs (considering multiple decades of demand and weather data) at greatly reduced computational cost, outperforming industry standard \textit{a priori} time series aggregation methods. As discussed in Section \ref{sec:introduction:demand_weather_uncertainty}, the consideration of such long samples is essential for robust power system planning under natural climate/weather variability. This method also allows the consideration of climate model data in power system planing (see e.g \cite{wohland_2017, craig_2019, van_zuijlen_2019}), where differences in climate have a weak signal-to-noise ratio under natural variability and hence require long samples to identify statistically significant effects. The models, data and sample code are available at \cite{github_iss}.

The key innovation behind the proposed methodology is a systematic identification and inclusion of relevant ``extreme'' events in subsamples through a user-defined \textit{importance} function. This function may be tailored to the specific application, but in many settings a time step's generation cost or electricity price will suffice (Section \ref{sec:importance_subsampling:intuition_remarks}). The systematic identification of relevant extremes may also be used in its own right, for example by using the identified ``extreme'' days for subsequent reliability analysis at higher spatiotemporal resolution.

\textit{Importance subampling} is an \textit{a posteriori} method (see Section \ref{sec:introduction:time_series_reduction}) in which properties of the model and/or its results are used to improve time series aggregation. The method can be compared with that in \cite{sun_2019}, where the time series aggregation is performed, along with a number of other modifications, across the vectors of model outputs (costs at technological and regional levels) for each individual day. This is another promising approach to a similar problem. \textit{Importance subsampling} is simpler, though a more thorough comparison of the two methods would be interesting.

Two potential extensions to the method are further refinements to model interdependence between the sampled days (as in e.g. weekly/seasonal storage), and dimensionality reduction when the number of grid locations becomes very large.

\bibliographystyle{IEEEtran}
\bibliography{IEEEabrv,citations}

\end{document}